\documentclass[12pt]{article}
\usepackage{graphicx,psfrag,epsf}

\newcommand{\bff}[1]{\mbox{\boldmath ${#1}$}}

\newcommand{\be}{\begin{equation}}
\newcommand{\ee}{\end{equation}}
\newcommand{\bea}{\begin{eqnarray}}
\newcommand{\eea}{\end{eqnarray}}

\def\beq{\begin{eqnarray}}
\def\eeq{\end{eqnarray}}

\def\be{\begin{equation}}
\def\ee{\end{equation}}

\begin{document}

\renewcommand{\thefootnote}{\fnsymbol{footnote}}
\begin{titlepage}

\begin{flushright}
PITHA 05/01\\
hep-ph/0501289\\[0.0cm]
31 January 2005
\end{flushright}

\vspace{1cm}
\begin{center}
\Large\bf\boldmath
Third-order Coulomb corrections\\ to the S-wave Green function, 
energy levels and wave functions at the origin
\end{center}

\vspace{0.5cm}
\begin{center}
M.~Beneke, 
Y. Kiyo, 
K. Schuller\\[0.3cm]
{\sl Institut f\"ur Theoretische Physik E, RWTH Aachen,\\ 
D -- 52056 Aachen, Germany
}
\end{center}

\vspace{0.8cm}
\begin{abstract}
\vspace{0.2cm}\noindent 
We obtain analytic expressions for the third-order corrections 
due to the strong interaction Coulomb potential 
to the $S$-wave Green function, energy levels and wave functions 
at the origin for arbitrary principal quantum number $n$. 
Together with the known non-Coulomb correction this results 
in the complete spectrum of $S$-states up to order 
$\alpha_s^5$. The numerical impact of these corrections on 
the Upsilon spectrum 
and the top quark pair production cross section near threshold 
is estimated. 
\end{abstract}

\end{titlepage}

\section{Introduction}
\label{sec:intro}

Several years ago advances in non-relativistic effective 
theory made it possible to compute quarkonium properties 
at the next-to-next-to-leading order (NNLO). Since the perturbative 
approach assumes that the non-relativistic energy scale 
$E\sim m\alpha_s^2$ is larger than the strong interaction 
scale $\Lambda_{\rm QCD}$, these computations apply to the lowest bottomonium 
states and heavy quark current spectral functions near threshold 
such as those relevant to top quark pair production. NNLO results have 
been obtained for matching 
coefficients \cite{Beneke:1997jm,Czarnecki:1997vz,Schroder:1998vy}, energy 
levels \cite{Pineda:1997hz,Melnikov:1998ug,Penin:1998kx,Beneke:1999qg}, 
wave functions at the origin \cite{Melnikov:1998ug,Penin:1998kx,Beneke:1999qg}
and spectral functions~\cite{Penin:1998kx,Beneke:1999qg,Hoang:1998xf,Melnikov:1998pr,Hoang:2000yr}, principally for the $S$-wave states, which 
are the most important ones for applications \cite{Beneke:1999zr}.

It was observed that the perturbative corrections 
are almost always very large. Although the origin of these large 
corrections can sometimes be understood as being due to mass
renormalization \cite{Beneke:1998rk} or large logarithms, 
it is currently believed that a complete third  
(next-to-next-to-next-to-leading/NNNLO) order calculation is 
necessary to describe accurately even the case of top quark pair
production near threshold. That this is a difficult undertaking
is reflected by the fact that partial results at 
NNNLO exist for various quantities \cite{Brambilla:1999qa,Kniehl:1999ud,Brambilla:1999xj,Kiyo:2000fr,Hoang:2000fm,Kniehl:2002br,Kniehl:1999mx,Manohar:2000kr,Hoang:2003ns,Kniehl:2002yv,Hoang:2001mm}, but only the $S$-wave ground state 
energy is currently fully known at third order \cite{Penin:2002zv}, save 
for an unknown constant $a_3$ in the Coulomb potential. 
In this paper, we take one further step and compute the NNNLO corrections 
from the QCD Coulomb potential to the $S$-wave energy levels and  
wave function at the origin for arbitrary principal quantum 
number $n$, and to the $S$-wave Green function (spectral function) 
relevant to top quark pair production. Together with the known 
non-Coulomb energy level corrections \cite{Kniehl:2002br,Penin:2005eu} 
this determines the $S$-wave 
energy levels completely at third order for any $n$. The correction 
to the Green function and wave functions at the origin forms part of 
the complete third-order top quark pair production cross section. 
Another motivation for first concentrating on the Coulomb corrections 
is that the Schr\"odinger equation with the Coulomb potential can be 
solved numerically, and the result can be compared to the 
perturbative computation. This is no longer possible once the 
singular non-Coulomb potentials are included, in which case the 
perturbative approach is the only option. Comparing the perturbative 
approximation to the numerical solution allows us to estimate the 
convergence of the successive approximations. 

\section{Outline of the calculation}
\label{sec:outline}

The computation of Coulomb corrections can be phrased in the language 
of elementary quantum mechanics. We consider the Hamiltonian 
\begin{equation}
H=-\frac{{\bf \nabla}^{\,2}}{m} + V(\bff{r}),
\end{equation}
where $m$ denotes the heavy quark pole mass, and $V$ the Coulomb 
potential of the strong force. With 
\begin{equation}
V(\bff{r}) = \int\frac{d^3\bff{q}}{(2\pi)^3}\,
e^{-i \mbox{\boldmath\footnotesize $q$}\mbox{\boldmath\footnotesize $x$}}\,
\tilde V(\bff{q}),
\end{equation}
the potential reads, up to the fourth order in the expansion in the
strong coupling $\alpha_s$,
\begin{eqnarray}
\tilde V(\bff{q}) &=& -\frac{4\pi C_F\alpha_s}{\bff{q}^2}+\delta 
\tilde V(\bff{q}), 
\\
\delta\tilde V(\bff{q}) &=& -\frac{4\pi C_F\alpha_s}{\bff{q}^2} 
\Bigg\{\frac{\alpha_s}{4\pi} \bigg[a_1+\beta_0\ln\frac{\mu^2}{\bff{q}^2}\bigg] 
+\left(\frac{\alpha_s}{4\pi}\right)^{\!2} \bigg[a_2 + 
\Big(2 a_1\beta_0+\beta_1\Big) \ln\frac{\mu^2}{\bff{q}^2} + 
\beta_0^2 \ln^2\frac{\mu^2}{\bff{q}^2}\bigg]
\nonumber\\
&&\,+\left(\frac{\alpha_s}{4\pi}\right)^{\!3} \bigg[a_3 +
8\pi^2 C_A^3\ln\frac{\nu^2}{\bff{q}^2} + 
\Big(3 a_2\beta_0+2 a_1\beta_1+\beta_2\Big)\ln\frac{\mu^2}{\bff{q}^2}
\nonumber\\
&&\hspace*{1.9cm}+ \,\Big(3 a_1\beta_0^2+\frac{5}{2}\beta_0\beta_1\Big)
\ln^2\frac{\mu^2}{\bff{q}^2} + 
\beta_0^3 \ln^3\frac{\mu^2}{\bff{q}^2}\bigg]
\Bigg\}.
\label{fullCoulomb}
\end{eqnarray}
Here $\beta_i$ are the coefficients of the QCD $\beta$-function 
in the $\overline{\rm MS}$-scheme,
defined with the convention $\partial a_s/\partial \ln\mu^2 = 
-\sum\beta_n a_s^{n+2}$, $a_s\equiv \alpha_s/(4\pi)$ such that 
$\beta_0 = 11 C_A/3-4 T_F n_f/3$ \cite{vanRitbergen:1997va}. 
The constants $a_1$, $a_2$ are given in \cite{Schroder:1998vy}, the 
$\alpha_s^4 C_A^3 \ln \nu^2/\bff{q}^2$ term is from
\cite{Brambilla:1999qa}, while the three-loop constant term $a_3$, 
currently unknown, is estimated to be 6240 ($n_f=4$) 
and 3840 ($n_f=5$) \cite{Chishtie:2001yb}. The group 
theory factors for SU($N_c$) gauge theory ($N_c=3$ in QCD) are 
$C_F=(N_c^2-1)/(2 N_c)$, $C_A=N_c$, $T_F=1/2$, and $n_f$ denotes the 
number of quarks whose masses are smaller than $m\alpha_s$ and 
neglected ($n_f=4$ for bottom systems, $n_f=5$ for top). $\alpha_s$ 
refers to the strong coupling in the $\overline{\rm MS}$ scheme at 
the renormalization scale $\mu$, while $\nu$ is a scale necessary to 
define the separation of potential and ultrasoft effects. The
dependence on this factorization scale cancels in physical quantities
when the ultrasoft correction is included. 

When $\delta\tilde V(\bff{q})=0$ the spectrum of the Hamiltonian 
reproduces the well-known Bohr energy levels. This will be our 
zeroth order approximation. We consider quarkonium systems in which 
$\alpha_s(m\alpha_s)$ is small, and compute the spectrum in a
perturbation expansion in $\alpha_s$. Hence, the coefficient of 
$\alpha_s^{n+1}$ in $\delta\tilde V(\bff{q})$ is considered a 
perturbation of the $n$th order, and the third-order result we are
aiming at requires up to three insertions of the $\alpha_s^2$ 
potential, but only one insertion of the order $\alpha_s^4$
potential. We shall focus on the $S$-wave Green function 
\begin{equation}
\label{GE}
G(E) \equiv \langle 0|\hat{G}(E) |0\rangle = 
\langle 0|\frac{1}{H-E-i\epsilon}|0\rangle,
\end{equation}
where $|0\rangle$ denotes a position eigenstate with eigenvalue 
$\bff{r}=0$, and compute the matrix element of the right-hand side of 
(omitting the argument $E$ of the Green function)
\begin{eqnarray}
\hat{G} &=& \hat{G}_0 - \hat{G}_0\delta V_1 \hat{G}_0- 
 \hat{G}_0\delta V_2 \hat{G}_0
 + \hat{G}_0\delta V_1 \hat{G}_0\delta V_1 \hat{G}_0
 \nonumber\\
 && - \,\hat{G}_0\delta V_3 \hat{G}_0+ 
 2\hat{G}_0\delta V_1 \hat{G}_0\delta V_2 \hat{G}_0
 - \hat{G}_0\delta V_1\hat{G}_0\delta V_1\hat{G}_0\delta V_1
   \hat{G}_0.
\label{expandedGreen}
\end{eqnarray}
$G_0(E)$ denotes the zeroth-order Green function, and $\delta V_n$ 
the $n$th order perturbation potential. The Green function $G(E)$ has
single poles at the exact $S$-wave energy levels $E=E_n$, such 
that 
\begin{equation}
G(E) \stackrel{E\to E_n}{=}\frac{|\psi_n(0)|^2}{E_n-E-i\epsilon}.
\label{nearpole}
\end{equation}
Inserting 
\begin{eqnarray}
E_n &=& E_n^{(0)} \left(1+\frac{\alpha_s}{4\pi} e_1 
+\left(\frac{\alpha_s}{4\pi}\right)^{\!2} e_2 
+\left(\frac{\alpha_s}{4\pi}\right)^{\!3} e_3 + \ldots\right),
\label{energyexpansion}
\\\
|\psi_n(0)|^2 &=& |\psi_n^{(0)}(0)|^2 \left(1+\frac{\alpha_s}{4\pi} f_1 
+\left(\frac{\alpha_s}{4\pi}\right)^{\!2} f_2 
+\left(\frac{\alpha_s}{4\pi}\right)^{\!3} f_3 + \ldots\right)
\label{psiexpansion}
\end{eqnarray}
into this equation, we can determine $e_{1,2,3}$ and $f_{1,2,3}$ 
by comparing the expansion of (\ref{nearpole}) 
in $\alpha_s$ with (\ref{expandedGreen}) 
near $E=E_n$. We recall that 
$E_n^{(0)} = -m (\alpha_s C_F)^2/(4 n^2)$, and 
$|\psi_n^{(0)}(0)|^2 = (m \alpha_s C_F)^3/(8\pi n^3)$.
The details of the calculation are too lengthy to be reproduced here. 
However, we sketch the computation of the threefold insertion 
$\langle 0|\hat{G}_0\delta V_1\hat{G}_0\delta V_1\hat{G}_0\delta V_1
\hat{G}_0|0\rangle$
of $\delta V_1$ in the Appendix.

There is an equivalent quantum-field-theoretical description of the 
calculation, which is the appropriate one in the context of systematic
higher-order computations of quarkonium properties
\cite{Beneke:1999zr}. 
After all the quantum-mechanical description breaks down beyond some 
accuracy, since (a) the system is sensitive to the short-distance quantum 
fluctuations, and (b) the restriction to the quark-antiquark sector 
of the Fock space is inadequate. Systematic calculations therefore 
start from QCD and obtain effective Lagrangians by systematically 
integrating out short-distance fluctuations ($\Delta x\sim 1/m$) 
as well as all massless degrees of freedom down to the ultrasoft 
energy scale $E\sim m\alpha_s^2$, which characterizes the size of 
energy fluctuations in a quarkonium bound state. The result is an effective 
Lagrangian, in which the potentials appear as matching 
coefficients \cite{Pineda:1997bj,Beneke:1998jj}. In particular 
the Coulomb potential is regarded as 
the matching coefficient of a relevant operator, and differs from the 
Wilson loop definition beginning at order $\alpha_s^4$. Consequently, 
the $\alpha_s^4\ln\alpha_s$ term which appears in the static 
potential \cite{Appelquist:1977es} is absent, and replaced by 
a logarithm of the factorization scale $\nu$ \cite{Brambilla:1999qa} 
as shown in (\ref{fullCoulomb}). The computation of 
Coulomb corrections is based on the Lagrangian 
\begin{eqnarray}
\label{pnrqcd}
{\cal L}_{\rm eff} &=& 
\psi^\dagger(x) \left(i \partial^0+
\frac{\bff{\partial}^2}{2m} \right)\psi(x) + 
\chi^\dagger(x) \left(i \partial^0-
\frac{\bff{\partial}^2}{2m} \right)\chi(x)  
\nonumber\\[0.0cm]
&& +\,\int d^3\bff{r}\,\left[\psi^\dagger \psi\right]\!(x+\bff{r}\,)\,
V(\bff{r})\left[\chi^\dagger \chi\right]\!(x),
\end{eqnarray}
where $V(\bff{r})$ is the (colour-singlet) 
Coulomb potential, and where the non-relativistic 
two-component 
field $\psi(x)$ ($\chi(x)$) destroys (creates) a heavy quark 
(anti-quark). There are other terms in the effective Lagrangian that 
have to be included for a complete NNNLO calculation. The point we
wish to make here is that these are well-understood, and hence there 
is a well-defined procedure to separate the complete calculation into
several simpler parts, of which the calculation of Coulomb corrections
is one. An important feature of (\ref{pnrqcd}) is that the
leading-order Coulomb potential cannot be treated as a perturbation, and 
must be included in the zeroth order approximation together with the 
bilinear terms. $\hat G_0$ is essentially the propagator of this 
theory. We can also obtain $G(E)$ in (\ref{GE}) directly by computing 
the two point function 
\begin{equation}
\label{nrcorr}
N_cG(E) = i \int d^4x\,e^{i q x}\, 
\langle\Omega|T([\psi^\dagger\chi](x) [\chi^\dagger\psi](0)|\Omega\rangle,
\end{equation}
with $|\Omega\rangle$ the Fock space vacuum state. 
This makes it clear that the Green function is closely related to  
inclusive heavy quark-anti-quark cross sections in $e^- e^+$
collisions, where the case of the  
top quark is particularly interesting.

\section{Third-order corrections to bound state parameters}
\label{sec:result}

In this section we present our result for the third-order 
correction from the Coulomb potential to the $S$-wave energy levels 
and wave functions at the origin (squared) for arbitrary 
principal quantum number $n$. The corrections (notation as in 
(\ref{energyexpansion},\ref{psiexpansion})) are parameterized as 
\begin{eqnarray}
&& e_{i}=e_{i}^{C}+e_{i}^{nC},
\\
&& f_{i}=f_{i}^{C}+f_{i}^{nC},
\end{eqnarray}
where $C$ stands for the correction, when only the Coulomb potential 
is included in the effective Lagrangian. By definition `$nC$' denotes 
all the remaining corrections, which originate from additional potentials 
and an ultrasoft non-potential interaction. 
Together with the non-Coulomb third-order correction $e_3^{nC}$ 
to energy level \cite{Kniehl:2002br,Penin:2005eu}, we obtain a 
complete result for the bound state energies to order $m\alpha_s^5$
for any $n$. For the wave function, however, $f_3^{nC}$ is not yet
known. We should emphasize that we define $|\psi_n(0)|^2$ by the 
residue of the correlation function (\ref{nrcorr}) of non-relativistic
currents. The non-Coulomb corrections arise from potentials which 
cause short-distance singularities, such that $f_i^{nC}$ are
factorization scheme-dependent quantities. This scheme-dependence 
is canceled by short-distance coefficients, which we do not discuss
here, but which are known to the same order as $f_i^{nC}$ is known 
($i\leq 2$).

\subsection{Energy levels}
The energy corrections from the Coulomb potential
are organized as follows,
\begin{eqnarray}
e_1^C &=&  4 \beta_0 \, L +  c_{E,1},
\\ 
e_2^C&=&      12\beta_0^2\, L^2 
            + L \,\bigg(-8 \beta_0^2 + 4 {\beta_1} 
                      + 6 \beta_0 c_{E,1}\bigg) 
            + c_{E,2} ,
\\
e_3^C&=&  32 \beta_0^3 \, L^3 
          +  L^2\, \bigg( - 56 \beta_0^3 
                          + 28 \beta_0 {\beta_1} 
                          + 24 \beta_0^2 c_{E,1}
                   \bigg) 
\nonumber \\ 
&&
              + L\, \bigg( 16 \beta_0^3 
                        - 16 \beta_0 {\beta_1} 
                        + 4 {\beta_2} 
                        - 12 \beta_0^2 c_{E,1} 
                        + 6 {\beta_1} c_{E,1} 
                        + 8 \beta_0 c_{E,2}
                   \bigg)
\nonumber \\
&&         
+ c_{E,3} 
+ 32 \pi^2 C_A^3 
   \bigg[\ln\left(\frac{n\nu}{m C_F \alpha_s}\right)+S_1(n)
   \bigg],
\end{eqnarray}
where $L=\ln\left(n\mu/(m C_F \alpha_s)\right)$ and 
$S_a(n)=\sum_{k=1}^n 1/k^a$
is the harmonic sum. For later convenience we introduce the nested harmonic
sums
\begin{eqnarray}
&& 
S_{a,b}(n)
\equiv \sum_{k=1}^n \frac{1}{ {k}^a }\, S_b(k),
\hspace{1cm}
S_{a,b,c}(n)\equiv \sum_{k=1}^{n}\frac{1}{k^a} \, S_{b,c}(k).
\end{eqnarray}
In the following we omit the argument of harmonic sums which is always
understood to be the principal quantum number $n$. 
The coefficients of the logarithmic terms are fixed by the 
renormalization group in terms of the $\beta$-function and the
$c_{E,i}$ from lower orders. The 
``non-trivial'' information is encoded in the constants $c_{E,\,i}$.
The first and second order corrections, $c_{E,\,1}, c_{E,\,2}$ are
known \cite{Pineda:1997hz,Melnikov:1998ug,Penin:1998kx,Beneke:1999qg}
\begin{eqnarray}
c_{E,\,1}&=& 
2 a_1 + 4 S_1 \beta_0,
\\
c_{E,\,2}&=& 
  a_1^2 
+ 2 a_2 
+ 4 S_1 \beta_1 
+ 4 a_1 \beta_0\bigg[\, 3 S_1-1 \,\bigg]  
\nonumber \\
&&
+ \beta_0^2 
     \Bigg[\, S_1 \bigg(12 S_1-8 - \frac{8}{ n}\bigg) 
          + 16 S_2 
          - 8 n S_3 
          + \frac{2 \pi^2}{ 3} 
          + 8 n {\xi(3)}
     \,\Bigg],
\end{eqnarray}
where $\xi(s)$ is the zeta-function, $\xi(s)=\sum_{k=1}^\infty k^{-s}$. 
Our new result is the third-order correction to $E_n$ for arbitrary $n$, 
which reads
\begin{eqnarray}
c_{E,\,3} &=& 
 2 a_1 a_2 
+ 2 a_3 
+ 2 a_1^2 \beta_0\Big[\, 4 S_1-5 \,\Big] 
+ 4 a_2 \beta_0  \Big[\, 4 S_1-1 \,\Big] 
+ 4 a_1 \beta_1  \Big[\, 3 S_1-1 \,\Big] 
\nonumber \\
&&
+ 4 S_1 \beta_2 
+ \beta_0 \beta_1 
      \Bigg[\,  
              S_1 \bigg(28 S_1-16 - \frac{24}{ n}\bigg) 
            + 36 S_2 
            - 16 n S_3
            +  \frac{7 \pi^2}{3}
            + 16 n {\xi(3)}
    \,\Bigg] 
\nonumber \\
&&
+ a_1 \beta_0^2 
      \Bigg[\, S_1 \bigg( 48 S_1-56 - \frac{32}{ n} \bigg) 
         + 64 S_2 
         - 32 n S_3 
         + 8 
         + \frac{8 \pi^2}{3} 
         + 32 n {\xi(3)}
     \, \Bigg]
\nonumber \\
&&
+ \beta_0^3 
     \Bigg[\, S_1 \Bigg( 
                     S_1 \bigg( 32 S_1 -56 - \frac{32}{ n}\bigg)  
                    + 96 S_2 
                    - 64 n S_3 
                    + 16 
                    + \frac{16}{n}
                    + \frac{32 \pi^2}{ 3} 
                    + 64 n {\xi(3)}
                  \Bigg) 
\nonumber \\
&&
             + S_2 \Bigg( 8 n S_2 
                        + 16 n^2 S_3
                        -32 
                        - \frac{16}{ n} 
                        - \frac{40 n \pi^2}{ 3} 
                        - 16 n^2 {\xi(3)}
                   \Bigg)
             + S_3 \Bigg(96 + 16 n + 8 n^2 \pi^2\Bigg)  
\nonumber \\
&&
            - 104 n S_4 
            + 48 n^2 S_5 
            - 144 S_{2, 1} 
            + 224 n S_{3, 1} 
            - 32 n^2 S_{3, 2} 
            - 96 n^2 S_{4, 1} 
            -\frac{4 \pi^2}{3} 
            + \frac{2 n \pi^4}{45} 
\nonumber \\
&&
            +{\xi(3)} \bigg(32 - 16 n - 8 n^2 \pi^2\bigg) 
            + 96 n^2 {\xi(5)}
      \,\Bigg].
\end{eqnarray}

For completeness we also give the non-Coulomb correction. 
The first-order correction is only from the Coulomb potential, 
thus $e_1^{nC}=0$. The 
second-order term was first obtained in \cite{Pineda:1997hz}, 
the third-order term for arbitrary $n$ in \cite{Penin:2005eu}. 
The expressions are
\begin{eqnarray}
e_2^{nC}/(16\pi^2) &=&
\frac{C_A C_F}{n}
+
\frac{{C_F}^2}{n}
      \left(2-\frac{11}{16n}-\frac{2 {\vec{S}}^{\,2} }{3}
      \right),
\\
e_3^{nC}/(64 \pi^2)
&=&
- \frac{49 n_f T_F C_A C_F}{ 36 n}
+
\frac{4 {C_A}^2 C_F}{3 n} 
\Bigg[  \frac{97}{48} 
      - \ln 2
      + \ln n
      + \ln(C_F\alpha_s)
      - S_1
\Bigg] 
\nonumber \\
&& \hspace*{-2cm}
+ 
\frac{{C_A}^3}{6} 
\Bigg[-\frac{5}{6} 
      - \ln 2
     - 2\ln n
     + 4\ln\left(C_F \alpha_s \right)
     - 3\ln(\nu/m)
     - 2 S_1
\Bigg] 
\nonumber \\
&&\hspace*{-2cm}
+ 
 C_A {C_F}^2 
\Bigg[
\frac{{\vec{S}}^{\,2} }{n}
 \left( - \frac{107}{ 108} 
          + \frac{7 \ln n}{ 6} 
          - \frac{7  S_1}{ 6} 
          - \frac{7 \ln(C_F\alpha_s)}{ 6} 
          + \frac{7}{ 12 n} 
    \right)
+
\frac{1}{n} 
\bigg(
            \frac{139}{ 36} 
          + \frac{7 \ln n}{ 6} 
\nonumber \\
&& \hspace*{-2cm}
          + \frac{41 \ln(C_F\alpha_s)}{ 6} 
          - \frac{7 S_1}{6} 
          - 4 \ln 2
\bigg)
          + \frac{2}{3 n^2}
            \left( -\frac{47}{16} 
                   + \ln 2
                   - \ln n
                   - \ln(C_F\alpha_s)
                   - S_1
             \right)
\Bigg]
\nonumber \\
&&\hspace*{-2cm}
+ 
\frac{{C_F}^3}{3 n} 
\Bigg[
      -2 n L_E(n) 
      + {\vec{S}}^{\, 2}
      -\frac{79}{6} 
      + \frac{7}{2 n} 
      - 8 \ln 2
      + 7 \ln n
      - 7 S_1
      + 9 \ln\left(C_F\alpha_s\right)
\Bigg] 
\nonumber \\
&&\hspace*{-2cm}
+\frac{ {C_F}^2 n_f T_F}{9 n} 
\Bigg[ -8
         + \frac{5}{ 2 n} 
         + \frac{10 \,{\vec{S}}^{\,2} }{3} 
\Bigg] 
+
\frac{T_F {C_F}^2}{ n}
\Bigg[
      \frac{32 }{ 15} 
      - 2 \ln 2 
      + {\vec{S}}^{\,2} \Big(\ln 2-1\Big)
\Bigg]
\nonumber \\
&&\hspace*{-2cm}
+ 
\beta_0
\Bigg[
C_A C_F 
  \bigg( \frac{2 }{ n} L
        -\frac{\pi^2}{6}
        + \frac{1}{2 n} 
        +  S_2
   \bigg) 
+ 
{C_F}^2
\Bigg( 
      \bigg(-\frac{11}{8 n^2} 
            +\frac{4}{n} 
            - \frac{4 {\vec{S}}^{\, 2}}{ 3n}
      \bigg) L
\nonumber \\
&&\hspace*{-2cm}
    + {\vec{S}}^{\, 2} 
        \left(  \frac{\pi^2}{ 9} 
               +\frac{1}{ 2 n} 
               -\frac{1}{6n^2}
               -\frac{2 S_2}{3} 
         \right) 
    - \frac{11 S_1 }{ 8 n^2}
    +   2 S_2 
    + \frac{1}{n} 
    + \frac{3}{8\,n^2} 
    -\frac{\pi^2}{ 3} 
 \Bigg)
\Bigg] 
\nonumber \\
&&\hspace*{-2cm}
+
\frac{a_1 C_A C_F}{2n} 
+
\frac{C_F^2 a_1}{2n}
\Bigg[ -\frac{9}{ 8 n} 
      +  \frac{7 }{2}   
      - {\vec{S}}^{\, 2}
\Bigg],
\end{eqnarray}
where ${\vec{S}^{\, 2}}$ is the eigenvalue of the spin operator. For the 
spin-triplet (singlet) state ${\vec{S}^{\, 2}}=2$ (${\vec{S}^{\, 2}}=0$). 
(The Coulomb potential is spin-independent, hence the $e_i^C$ do not 
depend on ${\vec{S}^{\, 2}}$.) Furthermore $L_E(n)$ denotes 
the ``Bethe logarithm'', which must be evaluated numerically. For 
$n=1,2,\ldots$ we find
\begin{equation}
L_E(n) = \left(-81.5379, -37.671, -22.4818, -14.5326, -9.52642,
  -6.0222,\ldots\right).
\end{equation}
The first three numbers have been obtained in \cite{Kniehl:1999ud}. 
We have performed an independent calculation of the ultrasoft 
correction.

\subsection{Wave functions at the origin}
The wave function corrections $f_i$ from the Coulomb potential 
are given by 
\begin{eqnarray}
f_1^C
&=& 
 6 \beta_0\, L 
+ c_{\psi,1},
\\
f_2^C 
&=&       
   24 \beta_0^2 \, L^2 
+  L\,\bigg(- 12 \beta_0^2 
            + 6 {\beta_1} 
            + 8 \beta_0 c_{\psi,1}  \bigg)
+ c_{\psi,2},
\\
f_3^C
&=&   
  80 \beta_0^3 \, L^3 
+  L^2 \,\bigg( -108 \beta_0^3 
                + 54 \beta_0 {\beta_1} 
                +  40 \beta_0^2 c_{\psi,1}  \bigg)
\nonumber \\
&& \hspace{-1cm}
+   L\, \bigg( 24 \beta_0^3 
             - 24 \beta_0 {\beta_1} 
              + 6 {\beta_2} 
              - 16 \beta_0^2 c_{\psi,1} 
              + 8 {\beta_1} c_{\psi,1} 
              + 10 \beta_0 c_{\psi,2}
                    \bigg) 
+ c_{\psi,3} 
\nonumber \\
&&
+ 48 \pi^2 {C_A}^3 
  \bigg[\ln\left(\frac{n \nu}{m C_F \alpha_s}\right)
         +\frac{1}{3}\left(S_1+ 2 n S_2-1-\frac{n \pi^2}{3}\right)
  \bigg].
\end{eqnarray}
The first and second order corrections are 
known  \cite{Melnikov:1998ug,Beneke:1999qg}
\begin{eqnarray}
c_{\psi,\,1}&=&
 3 a_1 
+2 \beta_0 \Bigg[\,  S_1 +  2 n S_2-1 - \frac{ n \pi^2}{ 3} \,\Bigg], 
\\
c_{\psi,\,2}&=&
3 a_1^2 
+ 3 a_2 
+ 2 a_1\beta_0 \Bigg[\, 4 S_1+ 8 n S_2-7- \frac{4 n \pi^2}{ 3}\,\Bigg] 
+ 2  \beta_1 \Bigg[\, S_1 + 2 n S_2-1 -  \frac{ n \pi^2}{ 3} \,\Bigg] 
\nonumber \\
&& \hspace*{-1cm}
+ \beta_0^2 
  \Bigg[\,
           S_1 \Bigg( 8 S_1 
                     + 16 n S_2
                     -20 
                     - \frac{12}{ n} 
                     - \frac{8 n\pi^2}{ 3} 
                \Bigg) 
         + S_2 \Bigg(  4 n^2 S_2 
                     + 8 
                     - 8 n 
                     - \frac{4 n^2 \pi^2}{ 3} 
               \Bigg) 
\nonumber \\
&& \hspace*{-1cm}
         + 28 n S_3 
         - 20 n^2 S_4 
         - 24 n S_{2, 1} 
         + 16 n^2 S_{3, 1} 
         +  4 
         + \frac{(3+4 n)\pi^2}{ 3}
         + \frac{n^2 \pi^4}{ 9} 
         + 20 n {\xi(3)}
  \, \Bigg].
\end{eqnarray}
Our new result for the third-order correction to $|\psi_n(0)|^2$ 
for arbitrary $n$ is 
\begin{eqnarray}
c_{\psi,\,3}&=&
  a_1^3 
+ 6 a_1 a_2 
+ 3 a_3 
+ 10 a_1^2 \beta_0 
    \Bigg[\,   S_1 
           +  2 n S_2 
           -  \frac{31}{10} 
           -  \frac{ n \pi^2}{ 3}
  \,\Bigg] 
+ 10 a_2 \beta_0 
    \Bigg[\,  S_1 
           +   2 n S_2
\nonumber \\
&&\hspace{-1cm}
           -  \frac{8}{5} 
           -  \frac{n \pi^2}{ 3} 
    \,\Bigg] 
+ 8 a_1 \beta_1 
     \Bigg[\,   S_1 
           +  2 n S_2
           - \frac{7}{4} 
           - \frac{ n \pi^2}{ 3} 
    \, \Bigg] 
+ 2 \beta_2 
    \Bigg[\,  S_1 
          + 2 n S_2
          - 1 
          - \frac{ n \pi^2}{ 3}  
  \, \Bigg] 
\nonumber \\
&&\hspace{-1cm}
+ \beta_0 \beta_1 
\Bigg[\,        S_1 \Bigg( 22 S_1 
                         + 40 n S_2 
                         - 44 
                         - \frac{36}{ n} 
                         - \frac{20 n \pi^2}{ 3} 
                     \Bigg) 
                +S_2 \Bigg(  8 n^2 S_2
                            + 14 
                            - 16 n 
                            - \frac{8 n^2 \pi^2}{ 3} 
                     \Bigg) 
\nonumber \\
&&\hspace{-1cm}
                + 64 n S_3 
                - 40 n^2 S_4 
                - 56 n S_{2, 1} 
                + 32 n^2 S_{3, 1} 
                + 8 
                +\frac{(21 + 16 n) \pi^2}{ 6} 
                + \frac{2 n^2 \pi^4}{ 9} 
                + 48 n {\xi(3)}
\,\Bigg]
\nonumber \\
&&\hspace{-1cm}
 +  a_1 \beta_0^2 
\Bigg[\,         S_1 \Bigg( 40 S_1 
                           + 80 n S_2
                           - 116 
                           - \frac{60}{ n} 
                           - \frac{40 n \pi^2}{ 3} 
                     \Bigg) 
               + S_2 \Bigg(   20 n^2 S_2
                            + 40 
                            - 72 n 
                            - \frac{20 n^2 \pi^2}{ 3} 
                     \Bigg)
\nonumber \\
&&\hspace{-1cm}
               + 140 n S_3 
               - 100 n^2 S_4 
               - 120 n S_{2, 1} 
               + 80 n^2 S_{3, 1} 
               + 48 
               + (5 + 12 n) \pi^2 
               + \frac{5 n^2 \pi^4}{ 9} 
               + 100 n {\xi(3)}
\Bigg] 
\nonumber \\
&&\hspace{-1cm}
+ \beta_0^3 
\Bigg[\,
       S_1 \Bigg(
               4 S_1 \Big( 4 S_1 
                        +  16 n S_2
                        -  19 
                        - \frac{6}{ n} 
                        - \frac{8 n \pi^2}{ 3} 
                     \Big) 
              + 8 S_2 \Big( 3 n^2 S_2
                        +  2 
                        -  14 n 
                        -  n^2 \pi^2 
                     \Big) 
\nonumber \\
&&\hspace{-1cm}
                + 104 n S_3 
              - 120 n^2 S_4
               - 112 n S_{2, 1} 
               + 96 n^2 S_{3, 1} 
               + 80 
               + \frac{64}{ n} 
               + \frac{(58 + 56 n) \pi^2}{ 3} 
               +\frac{2 n^2 \pi^4}{ 3} 
\nonumber \\
&&\hspace{-1cm}
               + 120 n {\xi(3)}
           \Bigg) 
     + S_2 \Bigg( -  4 n (17 +2 n) S_2 
                  +  72 n^2 S_3 
                  -  96 n^2 S_{2, 1} 
                  +  64 n^3 S_{3, 1} 
                  -  96 
                  + 16 n 
\nonumber \\
&&\hspace{-1cm}
                  - \frac{24}{ n} 
                  - \frac{8 (5 - n) n \pi^2}{ 3}
                  - 8 n^2 {\xi(3)}
           \Bigg) 
    +  S_3 \Bigg( - 16 n^3 S_3 +64 - 16 n - 20 n^2 \pi^2+ 32 n^3 {\xi(3)}
           \Bigg) 
\nonumber \\
&&\hspace{-1cm}
     + S_4 \Bigg(68 n + 40 n^2 + \frac{64 n^3 \pi^2}{ 3}
          \Bigg) 
     - 312 n^2 S_5 
     + 144 n^3 S_6 
     + S_{2, 1} \Bigg( 48 n 
                     - 120 
                     + 16 n^2 \pi^2 
               \Bigg) 
\nonumber \\
&&\hspace{-1cm}
     -32 S_{3, 1}\Bigg(\frac{15n }{2} 
                     +  n^2 
                     + \frac{n^3 \pi^2}{ 3}
               \Bigg) 
     + 384 n^2 S_{3, 2} 
     + 576 n^2 S_{4, 1} 
     - 224 n^3 S_{4, 2} 
     - 256 n^3 S_{5, 1} 
\nonumber \\
&&\hspace{-1cm}
     + 256 n S_{2, 1, 1} 
     + 64 n^2 S_{2, 2, 1} 
     - 64 n^3 S_{2, 3, 1} 
     - 448 n^2 S_{3, 1, 1} 
     + 192 n^3 S_{4, 1, 1} 
     - 8 
     - \frac{8(2 +  n) \pi^2}{ 3} 
\nonumber \\
&&\hspace{-1cm}
     - \frac{ (83 + 10 n)n \pi^4}{ 45} 
     + \frac{4 n^3 \pi^6}{ 105} 
     + {\xi(3)} \Bigg(48 - 80 n - 12 n^2 \pi^2 -16 n^3 {\xi(3)}\Bigg) 
     - 40 n^2 {\xi(5)}
\, \Bigg] \,.
\end{eqnarray}

The first-order non-Coulomb correction
vanishes, $f_1^{nC}=0$, as for the energy-levels. 
Starting from second order $|\psi_n(0)|^2$ 
depends on the factorization scheme that separates the hard
(relativistic) corrections from the low-energy corrections reproduced 
by the non-relativistic effective theory. The second-order term was 
first obtained in \cite{Melnikov:1998ug} for the spin-triplet 
states (${\vec{S}^{\, 2}}=2$), but the result given there does not 
refer to the conventional $\overline{\rm MS}$ subtraction scheme. 
The result of \cite{Melnikov:1998ug} including the hard correction 
has been reproduced in \cite{Beneke:1999qg}, where the $\overline{\rm
  MS}$ scheme was used for the individual contributions. Using these 
results (not printed in \cite{Beneke:1999qg}) we find for the $S$-wave
spin-triplet non-Coulomb correction to the wave function at the origin
squared in the 
$\overline{\rm MS}$ subtraction scheme 
\begin{eqnarray}
f_2^{nC}(\mu_h)|_{\rm \overline{MS}}/(16\pi^2)
&=&
C_F^2 \Bigg[
\frac{2}{3} L(\mu_h)-\frac{15}{8 n^2} +\frac{4}{3 n} + \frac{22}{9} 
-\frac{2}{3} S_1
\Bigg]
\nonumber\\
&& +\,
C_F C_A \Bigg[
 L(\mu_h)+\frac{2}{n}+\frac{5}{4}-S_1
\Bigg] \qquad ({\vec{S}^{\, 2}}=2),
\end{eqnarray}
where $L(\mu_f)=\ln\left(n\mu_f/(m C_F \alpha_s)\right)$ and $\mu_f$
refers to the factorization scale. The $\mu_f$ dependence cancels 
in the product $C^2\,|\psi_n(0)|^2$, where $C$ denotes the hard matching 
coefficient of the operator $\psi^\dagger\sigma^i\chi$
\cite{Beneke:1997jm,Czarnecki:1997vz}. 
($\sigma^i$ are the Pauli matrices.) 
The third-order coefficient $f_3^{nC}$ is unknown. 

\section{Quarkonium masses}
\label{sec:pheno}

In this section we compare the calculated energy levels 
with the bottomonium $\Upsilon(\rm{nS})$ masses. We also discuss 
the masses of the would-be toponium states, which are relevant 
to the top quark pair production cross section in 
high-energy $e^- e^+$ collisions. 

First we give a numerical version of the general result for the
$S$-state energy levels for the spin-triplet case 
($\vec{S}^{\,2}=2$), and $n_f=4$ (bottomonium) 
and $n_f=5$ (toponium). For the first three 
states $n=1,2,3$ we obtain, for $n_f=4$, 
\begin{eqnarray}
M_{\Upsilon(1S)}&=&2 m_b - \frac{4}{9} m_b  \alpha_s^2
\Bigg[1
+\bigg(3.590+2.653\,{L}\bigg)\alpha_s 
+\bigg(15.56+3.963_{nC}+12.07\, {L}
\nonumber \\
&&     +5.277\,{L}^2\bigg)\alpha_s^2
+\bigg(76.35+ 6.289\,\hat{a}_{3}
        + \big[28.47+15.30\,\ln\alpha_s+ 21.02\, {L}\big]_{nC}
\nonumber \\
&&      +72.65\, {L}+27.59\,{L}^2+9.332\,{L}^3\bigg)\alpha_s^3 \,\,
\Bigg]\, ,
\label{bottom1s}\\
M_{\Upsilon(2S)}&=&2 m_b - \frac{1}{9} m_b  \alpha_s^2
\Bigg[1
+\bigg(4.916+2.653\,{L}\bigg)\alpha_s
+\bigg(25.38+2.287_{nC}+17.34\,{L}
\nonumber \\
&&     +5.277\,{L}^2\bigg){\alpha_s}^2
+\bigg(140.7+6.289 \hat{a}_3 
       +\big[11.25+8.647\,\ln\alpha_s+12.13\,{L}\big]_{nC}
\nonumber \\
&&     +120.3\,{L}+41.59\,{L}^2+9.332\,{L}^3
 \bigg)\alpha_s^3 \,\,
\Bigg]\, ,
\\
M_{\Upsilon(3S)}&=&2 m_b - \frac{4}{81} m_b  \alpha_s^2
\Bigg[1
+\bigg(5.800+2.653\,{L}\bigg)\alpha_s
+\bigg(32.90+1.593_{nC}+20.86\,{L}
\nonumber \\
&&     +5.277\,{L}^2\bigg)\alpha_s^2
+\bigg(196.0+6.289\,\hat{a}_3 
       +\big[4.559+6.305\,\ln\alpha_s+8.449\,{L}\big]_{nC}
\nonumber \\
&&     +157.3\,{L}+50.92\,{L}^2+9.332\,{L}^3
\bigg)\alpha_s^3\,\,
\Bigg]\, .
\label{bottom3s}
\end{eqnarray}
Here $m_b$ denotes the bottom quark pole mass, 
$L=\ln(n\mu/(m_b C_F \alpha_s(\mu)))$, and we normalize the 
contribution from the unknown third-order constant in the Coulomb
potential, $a_3$, to the Pad\'{e} estimate by defining 
$\hat{a}_3=a_3/a_{3,\,Pade}$. We have given the Coulomb and 
non-Coulomb corrections separately to emphasize the numerical 
dominance of the former (in the pole scheme).
The quarkonium masses are of course independent 
of the ultrasoft factorization scale $\nu$, but the separation 
into a Coulomb and non-Coulomb correction is not. The representations 
of the series above is given for $\nu=m_b C_F \alpha_s(\mu)/n$. 
We note that the Coulomb correction increases with $n$, while the 
non-Coulomb corrections become smaller. The reason for this is 
that the characteristic distance scale (the ``Bohr radius'') 
increases $\langle r_n\rangle \sim n$, hence the relative effect of the 
short-range non-Coulomb interactions decreases for the excited 
states. The third-order result for $n=1$ has already been 
obtained in \cite{Penin:2002zv}, the other results are new. 

For the spin-triplet toponium, $n_f=5$, the series read 
\begin{eqnarray}
M_{t\bar{t}(1S)}&=&2 m_t-\frac{4}{9} m_t\alpha_s^2
\Bigg[1
+\bigg(3.201+2.440\,{L}\bigg)\alpha_s
+\bigg(12.47+3.963_{nC}+9.718\,{L}
\nonumber \\
&&
       +4.467\, {L}^2\bigg)\alpha_s^2
+\bigg(56.54+3.870\, \hat{a}_3
      +\Big[26.85+15.30\ln\alpha_s+19.34\,{L}\Big]_{nC}
\nonumber \\
&&     +52.88\, {L}+ 20.06\,{L}^2+7.267\, {L}^3
 \bigg) \alpha_s^3 \,    
\Bigg]\, ,
\label{top1s}\\
M_{t\bar{t}(2S)}&=&2 m_t-\frac{1}{9} m_t\alpha_s^2
\Bigg[1
+\bigg(4.421+2.440\,{L}\bigg)\alpha_s
+\bigg(20.54+2.287_{nC}+14.18\,{L}
\nonumber \\
&&
       +4.467\, {L}^2\bigg)\alpha_s^2
+\bigg(104.2+3.870 \, \hat{a}_3
      +\Big[10.35+8.647\,\ln\alpha_s+11.16\,{L}\Big]_{nC}
\nonumber \\
&&     +88.59\, {L}+ 30.96\,{L}^2+7.267\, {L}^3
 \bigg) \alpha_s^3 \,    
\Bigg]\, ,
\\
M_{t\bar{t}(3S)}&=&2 m_t-\frac{4}{81} m_t\alpha_s^2
\Bigg[1
+\bigg(5.234+2.440\,{L}\bigg)\alpha_s
+\bigg(26.74+1.593_{nC}+17.16\,{L}
\nonumber \\
&&
       +4.467\, {L}^2\bigg)\alpha_s^2
+\bigg(145.4+3.870\, \hat{a}_3
      +\Big[3.934+6.305\,\ln\alpha_s+7.773\,{L}\Big]_{nC}
\nonumber \\
&&     +116.4\,{L}+ 38.23\,{L}^2+7.267\, {L}^3
 \bigg) \alpha_s^3 \,    
\Bigg]\, ,
\end{eqnarray}
where $m_t$ is top quark pole mass. 

The series coefficients are large and the series do not converge 
for bottomonium, presumably because the pole mass introduces 
a strong infrared renormalon 
divergence \cite{Bigi:1994em,Beneke:1994sw}, which 
is not present in the physical observable ``quarkonium mass'' itself 
\cite{Beneke:1998rk,Hoang:1998nz}. It is therefore an advantage 
to use a better mass renormalization convention such as the 
potential-subtracted (PS) mass \cite{Beneke:1998rk}. The 
relation to the pole mass appropriate to third-order calculations 
is given by 
\begin{eqnarray}
m
&=&
m_{\rm PS}(\mu_f)
-\frac{1}{2}\int_{q \leq \mu_f}
 \frac{d^3 {\bff{q}}}{(2\pi)^3}\,
 \tilde{V}(\bff{q})|_{\nu=\mu_f}\, 
\nonumber \\
&=&
m_{\rm PS}(\mu_f)
+
\frac{\mu_f C_F\alpha_s}{\pi}
\Bigg[1 
+ \frac{\alpha_s}{4\pi} \Big(2 \beta_0\,l_1 +a_1\Big) 
+ \bigg(\frac{\alpha_s}{4\pi}\bigg)^2 
     \bigg( 4 \beta_0^2\,l_2+2\,\Big(2 a_1\beta_0+\beta_1\Big )l_1
     + a_2 \bigg) 
\nonumber \\
&& 
+ \bigg(\frac{\alpha_s}{4\pi}\bigg)^3 
  \bigg(   8 \beta_0^3 l_3 + 
   4 \Big(3 a_1\beta_0^2+\frac{5}{2}\beta_0\beta_1\Big) l_2 
   +2 \Big(3 a_2\beta_0+2 a_1\beta_1+\beta_2\Big) l_1 
\nonumber \\
&& \hspace{1.8cm}    
+ \,a_3 + 
   16 \pi^2\,C_A^3\bigg)
\Bigg],
 \label{PSmass}
\end{eqnarray}
where $l_1=\ln(\mu/\mu_f)+1$, $l_2=\ln^2(\mu/\mu_f)+2\ln(\mu/\mu_f)+2$,
$l_3=\ln^3(\mu/\mu_f)+3 \ln^2(\mu/\mu_f)+6\ln(\mu/\mu_f)+6$. 
Note that the PS mass is {\em defined} with the choice 
$\nu=\mu_f$ in the Coulomb potential $\tilde{V}(\bff{q})$. 
The scale $\mu_f$ should be chosen of order $m \alpha_s$ in order 
not to violate the power counting of the  
non-relativistic expansion, so the 
relation (\ref{PSmass}) is accurate to order $m \alpha_s^5$ just 
as the third-order bound state masses. Our standard choice is 
$\mu_f=2\,$GeV for bottom and $\mu_f=20\,$GeV for top. 


\subsection{Bottomonium}
Before extracting quark masses and predicting bottomonium masses 
it is instructive to display the convergence of the expansions 
at the ``natural'' renormalization scale $\mu$, where 
$L=\ln(n\mu/(M_b C_F\alpha_s(\mu)))=0$. (Here $M_b$ refers to the 
bottom quark mass in the chosen scheme.) 
For $\Lambda_{\rm QCD}^{(n_f=4)}=290.4\,$MeV, and with 4-loop running of 
$\alpha_s$ the ``natural'' scale is realized at  
$\mu=(2.02,\, 1.30,\, 1.03)\,$ GeV (for $n=1,2,3$) with
$m_b=5\,$GeV, and $\mu=(1.91, 1.23, 0.98)\,$ GeV 
with $m_{b,\rm PS}(2\,\mbox{GeV})=4.6\,$GeV. For the 
numerical analysis we adopt $a_3=a_{3,Pade}=6270$ ($\hat{a}_3=1$). 
Eqs.~(\ref{bottom1s}-\ref{bottom3s}) show that the precise value of 
$a_3$ is not important as long as it is not very different 
from the Pad\'{e} estimate.

Using the relation (\ref{PSmass}) between the pole and PS mass 
we re-express $M_{\Upsilon(nS)}$ in terms of $m_{b,\rm PS}$ 
taking into account that $\mu_f/m_b$ counts as 
one power of $\alpha_s$.  With $m_b=5 \,$GeV, 
$m_{b,\rm PS}\equiv m_{b,\rm PS}(2\,\mbox{GeV})=4.6\,$GeV we obtain 
\begin{eqnarray}
M_{\Upsilon(1S)}=2 m_b+E_1^{(0)}
&&\hspace{-3mm}
\bigg[
1+ 1.09_{\rm NLO} 
+\Big(1.42+0.36_{nC}\Big)_{\rm N^2LO}
+\Big(2.29+0.28_{nC}\Big)_{\rm N^3LO}
\bigg]
\nonumber \\
&& \hspace*{-3.6cm}= 2 m_{b,\rm PS}+E_{1,\rm PS}^{(0)}
\bigg[
1+ 0.19_{\rm NLO} 
+\Big(0.07-0.23_{nC}\Big)_{\rm N^2LO}
+\Big(0.09-0.19_{nC}\Big)_{\rm N^3LO}
\bigg],
\\
M_{\Upsilon(2S)}=2 m_b+E_2^{(0)}
&&\hspace{-3mm}
\bigg[
1+1.91_{\rm NLO}
+ \Big(3.84+0.35_{nC} \Big)_{\rm N^2LO}
+ \Big(8.64+0.18_{nC}\Big)_{\rm N^3LO} \,
\bigg] 
\nonumber \\
&& \hspace*{-3.6cm}=2 m_{b,\rm PS}+E_{2,\rm PS}^{(0)}
\bigg[
1+ 0.26_{\rm NLO} 
+\Big(0.26-0.05_{nC}\Big)_{\rm N^2LO}
+\Big(0.37-0.03_{nC}\Big)_{\rm N^3LO}
\bigg],
\label{mass2s}\\
M_{\Upsilon(3S)}=2 m_b+E_3^{(0)}
&&\hspace{-3mm}
\bigg[
 1
+2.69_{\rm NLO}
+ \Big(7.06+0.34_{nC} \Big)_{\rm N^2LO}
+ \Big(20.10-0.03_{nC}\Big)_{\rm N^3LO} \,
\bigg] 
\nonumber 
\\
&& \hspace*{-3.6cm}=2 m_{b,\rm PS}+E_{3,\rm PS}^{(0)}
\bigg[
1+ 0.25_{\rm NLO} 
+\Big(0.41-0.03_{nC}\Big)_{\rm N^2LO}
+\Big(0.69+0.00_{nC}\Big)_{\rm N^3LO}
\bigg]\,, 
\label{mass3s}
\end{eqnarray}
with 
\begin{equation}
E_{n,\rm PS}^{(0)}
=-\frac{ (\alpha_s C_F)^2 m_{b,\rm PS}}{4 n^2}
 +\frac{2\mu_f C_F\alpha_s}{\pi}.
\end{equation}
This clearly shows that the series expansions are useless in the 
pole scheme, but the successive terms are (slowly) decreasing in the PS scheme 
for $n=1$. For $n>1$ we still observe that the transition to the 
PS scheme eliminates the huge correction from the Coulomb potential
present in the pole scheme, yet the series coefficients are no longer 
converging. This is perhaps not surprising, because the scales 
are near or below $1\,$GeV for $n>1$, and a perturbative treatment is 
simply no longer justified. 

\begin{figure}[t]
\begin{center}
\makebox[0cm]{
\scalebox{0.7}{\rotatebox{0}{
     \includegraphics{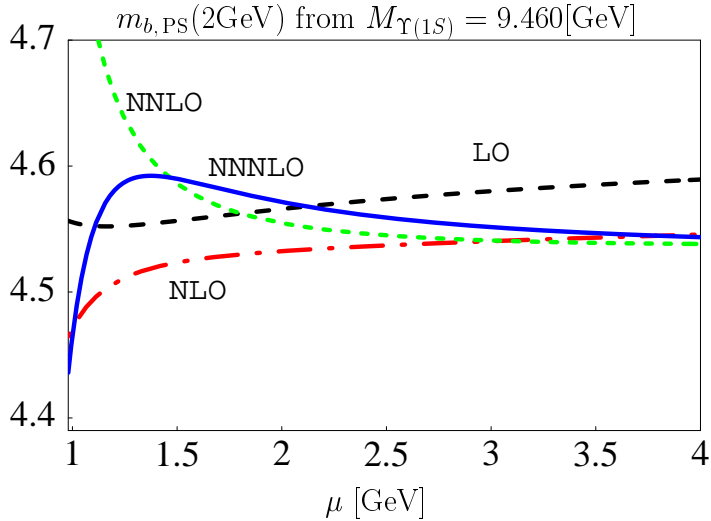}}}
}
\end{center}
\vspace*{-0.5cm}
\caption{
The bottom PS mass, $m_{b,\rm PS}(2\,\mbox{GeV})$, extracted from the 
experimental value $M_{\Upsilon(1S)}=9.460\,$GeV as a function of 
renormalization scale $\mu$ at LO (long dashes, black), 
NLO (long-short dashes, red), NNLO (short dashes, green) and 
NNNLO (solid, blue).} 
\label{fig:PSmass}
\end{figure}

We can therefore use the experimental $\Upsilon(\rm{1S})$ mass 
$M_{\Upsilon(1S)}=9.460\,$GeV to extract the bottom PS mass 
at NNNLO as was done in \cite{Beneke:1999fe} at NNLO. In 
Figure~\ref{fig:PSmass} we show the extracted PS mass as a 
function of the renormalization scale $\mu$ at LO (long dashes, black), 
NLO (long-short dashes, red), NNLO (short dashes, green) and 
NNNLO (solid, blue). Varying $\mu$ from $1.25\,$GeV to $4\,$GeV 
(as done in \cite{Beneke:1999fe}) the NNNLO correction is never larger 
than about $30\,$MeV and the error from the scale dependence is 
of similar size. We therefore assign a $\pm 30\,$MeV error to 
$m_{b,\rm PS}$ from the truncation of the perturbative expansion. 
The uncertainty in $\alpha_s(M_Z)=0.118\pm 0.003$ results in 
a $\pm 10\,$MeV error on $m_{b,\rm PS}$. The largest uncertainty 
in the determination of the bottom quark mass from the 
$\Upsilon(\rm{1S})$ mass is then from non-perturbative effects. 
The perturbative approximation is justified when the ultrasoft scale 
$m_b (C_F\alpha_s)^2\gg\Lambda_{\rm QCD}$, in which case the 
leading non-perturbative contributions is expressed in 
terms of the gluon condensate \cite{LV81,Vol82}
\begin{eqnarray}
&&\delta M_{\Upsilon(1S)}^{\rm np} = 
\frac{624\pi m_b }{425}\,
\frac{\langle\alpha_s GG\rangle}{(m_b C_F \alpha_s)^4}.
\end{eqnarray}
The numerical estimate is strongly dependent on the choice of scale 
in $\alpha_s$ in the denominator. Referring to \cite{Beneke:1999fe} 
for a more detailed discussion of the non-perturbative correction, 
we obtain 
\begin{equation}
m_{b,\rm PS}(2\,\mbox{GeV})
=(4.57 \pm 0.03_{\rm pert.}\pm 0.01_{\alpha_s} \pm 0.07_{\rm non-pert.})  
\,\mbox{GeV} 
\label{finalres}
\end{equation}
as the final result of this analysis. Because of the small 
third-order correction the bottom quark mass remains 
practically unchanged compared to the second-order analysis 
of \cite{Beneke:1999fe}, and so does the $\overline{\rm MS}$ 
mass determined from (\ref{finalres}). Further improvement of 
the mass determination by this method requires a quantitative 
understanding of non-perturbative effects. 

\begin{figure}
\begin{center}
\makebox[0cm]{
\hspace{-0.5cm}
\scalebox{0.55}{\rotatebox{0}{
     \includegraphics{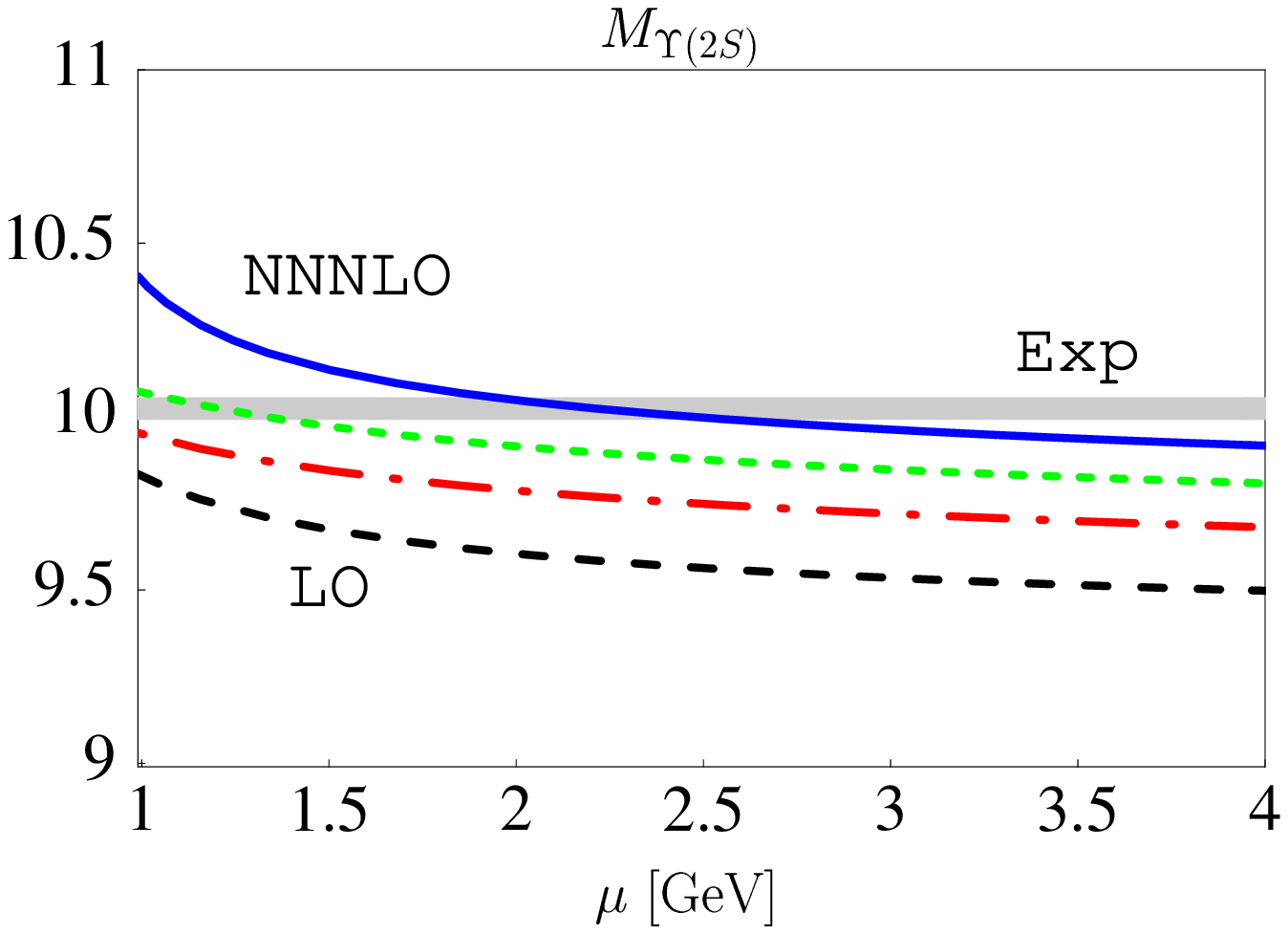}}}
\hspace{-0.5cm}
\scalebox{0.55}{\rotatebox{0}{
     \includegraphics{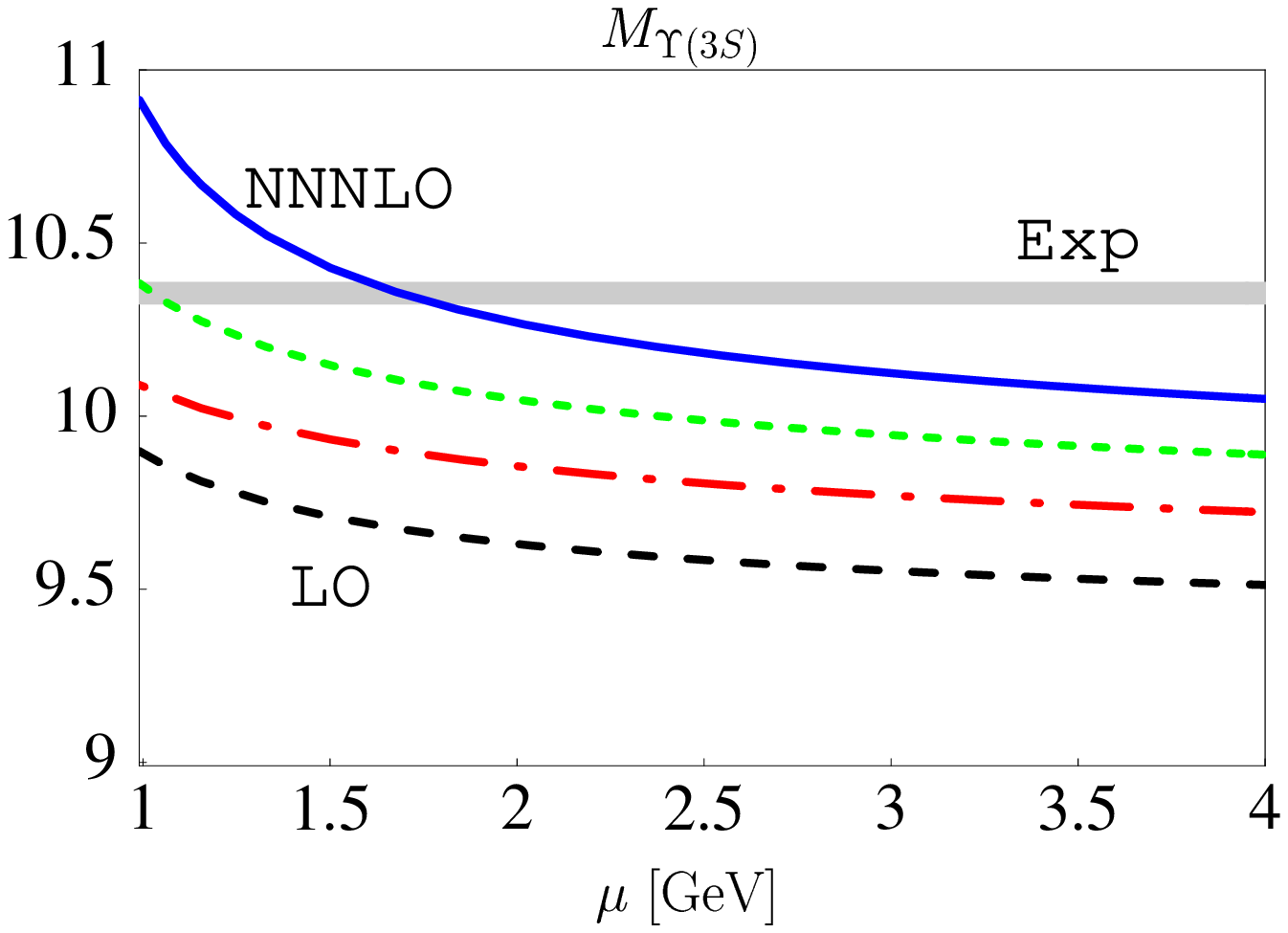}}}
}
\end{center}
\vspace*{-0.3cm}
\caption{\label{fig:2S3S}
Predicted masses of the $\Upsilon(\rm{2S})$ and $\Upsilon(\rm{3S})$ 
as a function of the renormalization scale $\mu$.  The lines 
refer to LO (long dashes, black), 
NLO (long-short dashes, red), NNLO (short dashes, green) and 
NNNLO (solid, blue). The widths of the bands for the experimental mass 
values are exaggerated.}
\end{figure}

Having determined $m_{b,\rm PS}$, we are in the position to predict 
the masses of the excited $S$-level states at the third order. 
(An analysis of the complete spectrum at second order was 
performed in \cite{Brambilla:2001fw}.) We focus on the 
spin-triplet states $\Upsilon(\rm{2S})$ and $\Upsilon(\rm{3S})$. 
The successive approximations up to the third order are shown in 
Figure~\ref{fig:2S3S} for $m_{b,\rm PS}=4.57\,$GeV as a function of 
the renormalization scale. For $\mu$ between $2\,$GeV and $4\,$GeV 
it appears that the large third-order correction is welcome to 
bring the theoretical result closer to the observed masses. However, 
at the natural scales $1.23\,$GeV ($n=2$) and $0.98\,$GeV ($n=3$) 
the NNLO result agrees well with data and the NNNLO correction 
renders the prediction too large. As is apparent from the figure, 
the conclusion is that the perturbative computation of bottomonium 
masses is applicable only to the ground state, $n=1$, while 
the excited states, involving larger distances, appear to be 
in the non-perturbative regime. It can be seen 
from (\ref{mass2s},\ref{mass3s}) that the 
NNNLO term for $n>1$ is dominated by the Coulomb correction. 

\subsection{Toponium}

We briefly discuss the situation for the toponium masses. Toponium 
bound states do not exist in nature due to large decay width 
$\Gamma_t\sim 1.5\,$GeV of the top quark \cite{Bigi:1986jk}, however 
the remnant 
of the 1S toponium state should be visible as an enhancement 
in the cross section $e^-e^+\to t\bar t X$ near threshold. The 
convergence of the series expansion for the toponium 1S mass is 
therefore a good measure for the accuracy to which the top 
quark mass might be determined from this cross section \cite{Hoang:2000yr}. 

The method suggested in \cite{Beneke:1999qg,Beneke:1998jj} relies 
on determining $m_{t,\rm PS}\equiv m_{t,\rm PS}(20\,\mbox{GeV})$ 
from the cross section measurement and obtaining the 
top quark $\overline{\rm MS}$ mass from $m_{t,\rm PS}$, since 
both relations are expected to be expressible in terms of 
well-behaved perturbative expansions. Adopting 
$m_{t,\rm PS}=175\,$GeV, $n_f=5$, 
$\Lambda_{\rm QCD}^{(n_f=5)}=208\,$MeV and the natural 
scale $\mu=32.6\,$GeV, where $L=0$, we obtain 
\begin{equation}
M_{t\bar t(1S)}=(350+0.85+0.05-0.13+0.01)\,\mbox{GeV} = 350.78\,\mbox{GeV}.
\end{equation}
The sum of the series varies only by about $60\,$MeV when the scale 
is taken between $15\,$ and $60\,$GeV, although the convergence is 
no longer satisfactory at the lower scale. The small uncertainty 
implies that  $m_{t,\rm PS}$ can be determined with little theoretical 
error from the cross section. The largest uncertainty in the 
determination of the top quark $\overline {\rm MS}$ mass then 
results from the unknown four-loop coefficient in the relation 
between the  $\overline {\rm MS}$ mass and the pole mass, which 
is needed to convert $m_{t,\rm PS}$ to the 
$\overline {\rm MS}$ mass, as already observed 
in \cite{Beneke:1999qg}. This uncertainty is estimated to be 
around $100\,$MeV (see Table~2 of \cite{Hoang:2000yr}). Our conclusions 
are in good agreement with \cite{Kiyo:2002rr}, where one of 
us investigated the direct determination of the top quark 
$\overline {\rm MS}$ mass from $M_{t\bar t(1S)}$ also using 
the NNNLO result for the 1S energy level. We should emphasize 
that none of these estimates take into account electroweak 
corrections, which are non-negligible, and must be included 
in the mass relations and cross section prediction before a 
comparison with the experimental cross section can be attempted.


\section{Third-order Coulomb wave functions at the origin and 
Green function}
\label{sec:green}

In this section we turn to the discussion of the $S$-wave 
Coulomb Green function, and to the wave function of the origin squared 
(residues of the Green function at the bound state poles). Since 
the third-order correction is not completely known, we include 
in this section only the Coulomb corrections as defined in 
Section~\ref{sec:result}, i.e.~we also neglect the (known) 
non-Coulomb correction at second order. The series expansions 
seem to be out of control for the wave functions in the bottomonium 
system, hence we focus on the case of the top quark and set 
$n_f=5$. We also set $a_3=a_{3,Pade}=3840$.

\subsection{Wave function at origin squared}
The  numerical version of the general result for the
$S$-state Coulomb wave function at the origin squared 
reads, for $n=1,2,3$, 
\begin{eqnarray}
\big|\psi_1(0)\big|^2_C
&=&
\frac{(m_t C_F\alpha_s)^3}{8\pi}
\bigg[
1
+\alpha_s \Big(-0.4333 + 3.661\,L\Big) 
+\alpha_s^2 \Big(5.832-5.112\,L + 8.933\,L^2\Big) 
\nonumber \\
&& \hspace*{-1cm}
+\,\alpha_s^3 \Big(-13.73
                 +6.446\,\ln\left(\frac{\nu}{m_t C_F \alpha_s}\right) 
                 + 39.72\,L- 22.91\,L^2 + 18.17\,L^3\Big)
\bigg],
\label{toppsi1}\\[0.2cm]
\big|\psi_2(0)\big|^2_C
&=&
\frac{(m_t C_F\alpha_s)^3}{64\pi}
\bigg[
1
+\alpha_s \Big(-0.1769 + 3.661\,L\Big) 
+\alpha_s^2 \Big(10.19-3.861\,L + 8.933\,L^2\Big) 
\nonumber \\
&&\hspace*{-1cm}
+\,\alpha_s^3 \Big(- 20.36
             + 6.446\,\ln\left(\frac{2\nu}{m_t C_F \alpha_s}\right) 
             + 65.31\,L - 19.09\,L^2+18.17\,L^3\Big)
\bigg],
\\[0.2cm]
\big|\psi_3(0)\big|^2_C
&=&
\frac{(m_t C_F\alpha_s)^3}{216\pi}
\bigg[
1
+\alpha_s \Big(0.07953+3.661\,L\Big) 
+\alpha_s^2 \Big(13.27-2.609\,L+ 8.933\,L^2\Big) 
\nonumber \\
&&\hspace*{-1cm}
+\,\alpha_s^3 \Big(-22.86
             +6.446\,\ln\left(\frac{3\nu}{m_t C_F \alpha_s}\right) 
             + 83.07\,L- 15.27\,L^2+ 18.17\,L^3\Big)
\bigg]\, .
\end{eqnarray}

\begin{figure}[t]
\begin{center}
\makebox[0cm]{
\scalebox{0.7}{\rotatebox{0}{
     \includegraphics{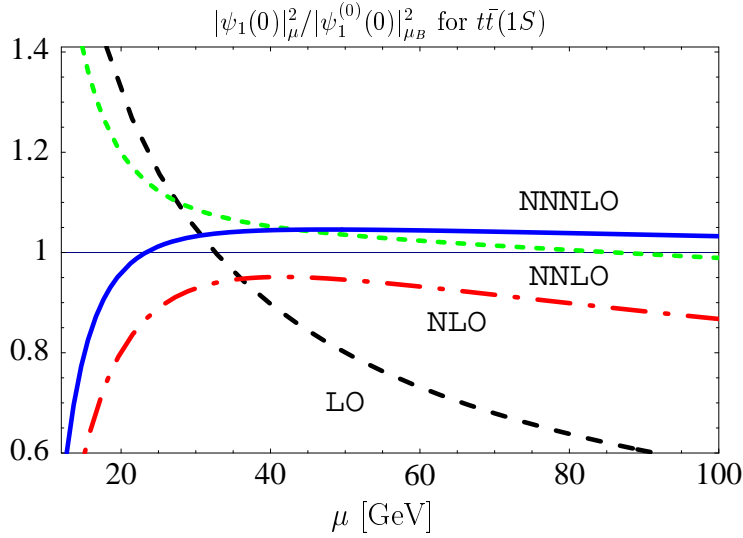}}}
}
\end{center}
\vspace*{-0.5cm}
\caption{The Coulomb wave function at the origin squared for 
the ground state ($n=1$) normalized
by $|\Psi_1^{(0)}(0)|^2$ at $\mu_B=32.6\,$GeV is shown as a function of 
the renormalization scale $\mu$. The 
input parameters are $m_{t,\rm PS}(20\,\mbox{GeV})=175\,$GeV, 
$\nu=m_{t,\rm PS} C_F \alpha_s(\mu)$. The lines 
refer to LO (long dashes, black), 
NLO (long-short dashes, red), NNLO (short dashes, green) and 
NNNLO (solid, blue).}
\label{fig:wf}
\end{figure}
We show the successive approximations for $n=1$ in 
Figure~\ref{fig:wf}. As before, we re-expressed the expansion 
in terms of the PS mass, treating  $\mu_f\sim m_t\alpha_s$. 
However, we note that contrary to the energy levels, the 
introduction of the PS mass does not change qualitatively 
the behaviour of the expansion, neither would this be 
expected on theoretical grounds. Our reference top quark mass 
is $m_{t,\rm PS}(20\,\mbox{GeV})=175\,$GeV, and the ultrasoft factorization 
scale is chosen to be $\nu=m_{t,\rm PS}C_F \alpha_s(\mu)$ such that the 
corresponding logarithm in (\ref{toppsi1}) vanishes. We also 
normalized the result to the LO wave function at the scale 
$\mu_B\equiv m_{t,\rm PS}C_F \alpha_s(\mu_B)=32.6\,$GeV. It is 
clearly seen from the figure that the approximations converge, and that 
the inclusion of the new 
third-order correction stabilizes the prediction further, 
provided $\mu$ is larger than about $25\,$GeV. We find a similar 
behaviour for $n=2,3$, where, however, the enhancement of the 
wave function relative to leading order is about $50\%$ ($n=2$) 
and $100\%$ ($n=3$) rather than roughly $5\%$ as at $n=1$. 

It may be disconcerting 
that the perturbative expansion breaks down already at 
scales as large as $20\,$GeV, where the strong coupling is still 
small. A more detailed analysis shows that this early breakdown 
is caused primarily by the $\alpha_s (\alpha_s\beta_0\ln\bff{q}^2)^n$ 
terms in the Coulomb potential. We also note that at $\mu=\mu_B$, 
where $L=0$, the series expansions shown above are sign-alternating 
in contrast to the corresponding expressions for the energy levels, 
which exhibit fixed-sign behaviour. It is a general fact that 
for sign-alternating series of a certain regularity, the convergence 
of the expansion is much improved by choosing a larger 
scale \cite{Beneke:1992ea}, since this renders the series coefficients 
{\em and} the expansion parameter $\alpha_s$ smaller. This 
explains the stability seen in the figure towards scales 
larger than the natural scale $\mu=\mu_B$, and suggests that 
an error estimate for $|\psi_n(0)|^2$ from varying the scale between $\mu_B/2$ 
and $2 \mu_B$ may be misleadingly large. We shall see in the 
following subsection that this is indeed the case. 

\subsection{Green function}

In addition to the energy levels and wave functions at the origin 
we have also computed the full $S$-wave 
Green function up to the third order in the presence of the 
Coulomb potential as described in Section~\ref{sec:outline}. 
The result is expressed in terms of multiple sums that can be 
evaluated numerically only. 

The Coulomb Green function plays an important role in the calculation 
of inclusive top quark pair production $e^- e^+\to t\bar t X$ near 
threshold \cite{FK87}, since the bulk of the cross section is given by 
\begin{equation}
R=\frac{\sigma_{t\overline{t}X}}{\sigma_{\mu^{+}\mu^{-}}}=
\frac{18\pi e^{2}_{t}}{m_t^2} (1+a_Z) \,\mathrm{Im}\,G(E+i\Gamma_t)
\end{equation}
where $e_t=2/3$ is the top quark electric charge, $\Gamma_t$
the top quark width, and $a_Z$ accounts 
for the vector coupling to the $Z$ boson. The convergence of the 
perturbative approximation up to the second order has been the 
subject of many investigations several years ago (see the 
review \cite{Hoang:2000yr}).

We are now in the position to extend this investigation to the third order 
as far as the corrections from the Coulomb potential are concerned. 
We computed 
the perturbative expansion of the Green function given 
in (\ref{expandedGreen}). This strict expansion is never a good 
approximation, because it contains terms of the form
\begin{equation}
\left[\frac{\alpha_s E_n^{(0)}}{E_n^{(0)}-(E+i\Gamma_t)}\right]^k,
\end{equation}
which originate from the expansion of (\ref{nearpole}) around 
$E_n^{(0)}$ rather than the true pole position. These terms become 
numerically large near $E\approx E_n^{(0)}$, but they can be summed 
by adding the exact pole structure and subtracting the expanded 
structure to the appropriate order, see \cite{Beneke:1999qg} for 
the corresponding expressions. In the following, ``perturbative 
approximation'' means that this resummation is included. 
Alternatively, we computed the Green function (\ref{GE}) 
numerically by solving the Schr\"odinger equation with the 
Coulomb potential (\ref{fullCoulomb}) exactly, following the 
method described in \cite{Strassler:1990nw}. We shall refer to this 
as the ``exact result''. The exact result contains an arbitrary 
number of insertions of the perturbation potentials 
$\delta V_1$, $\delta V_2$, $\delta V_3$. The leading difference to 
the third-order perturbative approximation consists of fourth-order 
terms of the form $\hat{G}_0\delta V_1\hat{G}_0\delta V_1\hat{G}_0\delta V_1
   \hat{G}_0\delta V_1\hat{G}_0$, 
$\hat{G}_0\delta V_1\hat{G}_0\delta V_1\hat{G}_0\delta V_2
   \hat{G}_0$ etc. Comparing the two results we obtain an estimate of 
the importance of these multiple insertions and of the convergence 
of the perturbative approximation. 

For the following numerical study we assume 
$\Lambda^{(n_f=5)}_{\rm QCD}=0.208\,$GeV ($\alpha_s(M_Z)=0.118$), four-loop 
evolution of $\alpha_s$, $\Gamma_t=1.5$GeV, and $\nu=20\,$GeV. 
It is crucial for an accurate 
prediction of the cross section not to use the 
top quark pole mass as an input 
parameter \cite{Beneke:1999qg,Hoang:2000yr}. Our result is 
presented in terms of the top-quark potential-subtracted 
mass $m_{t,\rm PS}(20\,\mbox{GeV})=175\,$GeV defined  
in (\ref{PSmass}). The conversion to the top quark $\overline{\rm MS}$ 
mass is given in \cite{Hoang:2000yr}. We implement the PS 
scheme by working with an order-dependent pole mass. That is, 
in the N${}^k$LO perturbative approximation we compute 
the top quark pole mass $m_t^{(k)}$ 
from  $m_{t,\rm PS}(20\,\mbox{GeV})=175\,$GeV 
according to (\ref{PSmass}) including the terms up to order 
$\mu_f\alpha^{k+1}_s$, and use $m_t^{(k)}$ as an input to the Green 
function. For instance, for renormalization scale $\mu=30\,$GeV we 
obtain $m_t^{(k)} = (176.21,176.56,176.74,176.87)\,$GeV for 
$k=0,1,2,3$. The energy argument of the Green function is then 
\begin{equation}
E=\sqrt{s}-2m_{t}^{(k)}
\end{equation}
with $\sqrt{s}$ the center-of-mass energy of the $e^- e^+$ collision. 

\begin{figure}[p]
\begin{center}
   \hskip-6cm
   \epsfxsize=18cm
   \centerline{\epsffile{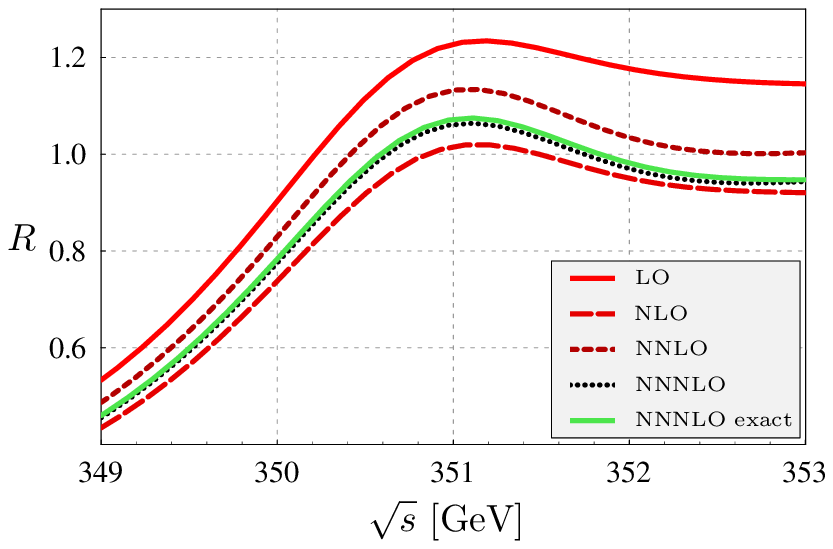}}
    \hspace*{-6cm}
   \epsfxsize=18cm
   \centerline{\epsffile{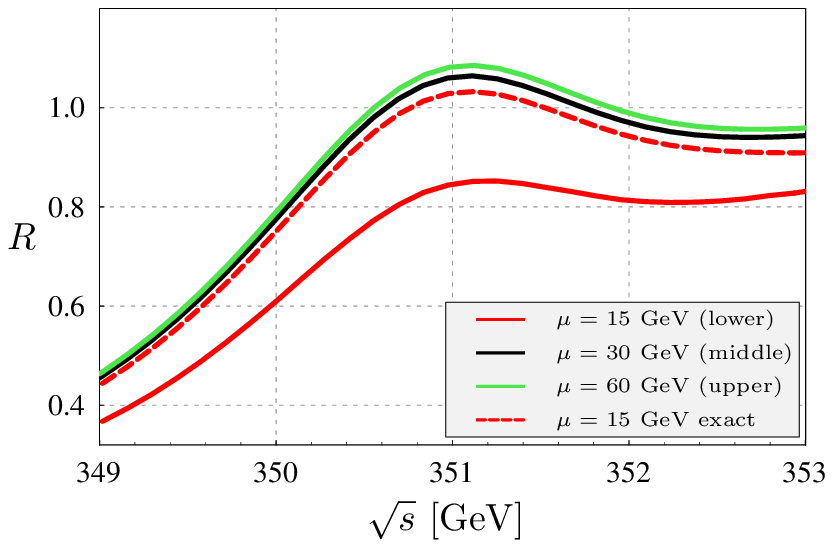}}
\caption{Top quark pair production cross section (Coulomb corrections 
only) for $m_{t,\rm PS}=175\,$GeV, $\Gamma_t=1.5\,$GeV. 
Upper panel: successive approximations 
up to the third order for $\mu=30\,$GeV. Lower panel: Scale dependence 
of the third-order approximation. See text for further explanation.}
  \label{fig:cs_order}
 \end{center}
\end{figure}

The upper plot in Figure~\ref{fig:cs_order} shows the convergence of the 
successive perturbative approximations to the exact result for the 
Green function (cross section) at the renormalization scale 
$\mu=30\,$GeV. The location of the ``peak'' position is indeed 
stable under the inclusion of higher-order corrections as expected 
in the PS scheme. On the other hand, the corrections to the magnitude 
of the cross section near the peak are significant, decreasing from  
about $20\%$ at NLO to about $7\%$ at NNNLO. The corrections alternate 
in sign as expected from the behaviour of the series expansion 
of $|\psi_1(0)|^2$. An important observation is that the third-order 
perturbative approximation coincides with the exact result 
within $1\%$. Hence the higher-order insertions of the perturbation 
potentials are negligible. The convergence of the approximations 
then suggests that the residual error from yet higher perturbation 
Coulomb potentials $\delta V_4$ etc. should be less than $5\%$. 

In the lower plot of Figure~\ref{fig:cs_order} we display the 
renormalization scale dependence of the third-order perturbative 
approximation (solid lines, $\mu=(60,30,15)\,$GeV from top 
to bottom). It is immediately apparent that the scale dependence 
is very small from $30$ to $60\,$GeV, but the result for 
$\mu=15\,$GeV is far away. We can trace this anomalous behaviour 
directly to the breakdown of the perturbative expansion for 
$|\psi_1(0)|^2$ at scales below $25\,$GeV discussed in 
the previous subsection and displayed in Figure~\ref{fig:wf}. 
The exact result (dashed line) does {\em not} exhibit this 
behaviour for $\mu=15\,$GeV, hence we conclude that the 
multiple insertions of the perturbation potentials become 
large at small scales and destroy the agreement of the perturbative 
and exact result. Indeed, we find that the series of multiple 
insertions is very slowly converging at small 
scales. We therefore learn the important lesson that the ``correct'' 
choice of scale in the perturbative approach is 
$\mu>25\,$GeV, while choosing smaller scales would lead to misleadingly 
large uncertainties. The lower plot of Figure~\ref{fig:cs_order} 
indicates that the scale dependence is less than $5\%$, similar in size 
to the truncation error estimated above. This discussion does not 
include the unknown third-order non-Coulomb corrections, but it 
lends support to the optimistic interpretation that the magnitude 
of the $t\bar t$ threshold cross section can eventually be 
computed with an accuracy of a few percent.

\section{Conclusion}
\label{sec:conclude}

We computed the third-order corrections from the strong interaction 
Coulomb potential to the energy levels and the wave functions at the 
origin of the $S$-waves bound states, and to the expectation value 
of the $S$-wave Green function operator in the $|\bff{r}=\bff{0}\rangle$ 
state. We view this as a first step towards a complete calculation 
of the third-order top quark pair production cross section 
$e^- e^+\to t\bar t X$ 
near threshold. It also completes the expression for the 
$S$-wave quarkonium masses with accuracy $m\alpha_s^5$ 
for arbitrary principal quantum 
number $n$, except 
for the unknown constant $a_3$ in the Coulomb potential. 

We updated the determination of the bottom quark mass from the 
mass of the $\Upsilon(\rm{1S})$ and obtain
\begin{equation}
m_{b,\rm PS}(2\,\mbox{GeV})
=(4.57 \pm 0.03_{\rm pert.}\pm 0.01_{\alpha_s} \pm 0.07_{\rm non-pert.})  
\,\mbox{GeV} 
\end{equation}
for the PS mass, with almost no modification compared to the 
second-order analysis \cite{Beneke:1999fe}. Our numerical study 
of the Coulomb corrections to the top quark pair production cross section 
near threshold shows that the perturbative approach (mandatory once 
non-Coulomb corrections are included) works and led to the 
conclusion that the residual uncertainty from the Coulomb 
corrections is less than $5\%$. This 
lends support to the optimistic interpretation that the magnitude 
of the $t\bar t$ threshold cross section can eventually be 
computed with an accuracy of a few percent. 

\vskip0.4cm\noindent
{\bf Note added}

\vskip0.2cm\noindent
During the preparation of this paper Penin, Smirnov and Steinhauser 
\cite{Penin:2005eu} have obtained 
results for the Coulomb correction to the second and third energy level 
and wave function at the origin ($n=2,3$), which agree with our 
result for general $n$. 
We thank the authors for communicating and 
comparing their results prior to publication. 

\vskip0.4cm\noindent
{\bf Acknowledgements}

\vskip0.2cm\noindent 
This work was supported by the DFG 
Sonderforschungsbereich/Transregio 9 ``Computer-gest\"utzte Theoretische 
Teilchenphysik''.  K.S. acknowledges support of the 
DFG Gra\-duier\-ten\-kolleg ``Elementarteilchenphysik an der TeV-Skala''.

\section*{Appendix}
\label{sec:app}

The most difficult part of the third-order calculation of the Green 
function is the threefold insertion 
$\langle 0|\hat{G}_0\delta V_1\hat{G}_0\delta V_1\hat{G}_0\delta V_1
\hat{G}_0|0\rangle$ of the first-order perturbation potential 
\begin{equation}
\delta\tilde V_1(\bff{q}) = -\frac{4\pi C_F\alpha_s}{\bff{q}^2} 
\frac{\alpha_s}{4\pi} \bigg[a_1+\beta_0\ln\frac{\mu^2}{\bff{q}^2}\bigg]. 
\end{equation}
The matrix element is ultraviolet and infrared finite, so the 
computation can be done in three dimensions. Since the external state 
is a position eigenstate, we Fourier transform to position space, 
where the potential has a $1/r$ and a $\ln(r)/r$ term. We can 
generate these terms by working with 
\begin{eqnarray}
W_i(\bff{r}_i)&=&\frac{1}{4\pi\Gamma(1+2u_i)\cos(\pi u_i)}
\left(\bff{r}_i^2\right)^{u_i-\frac{1}{2}}
\end{eqnarray}
and by taking the zeroth and first derivative with respect to the 
$u_i$ at $u_i=0$. The threefold insertion of the generating potential 
reads 
\begin{eqnarray}
J_{3}&\equiv& \int \prod_{i=1}^{3} d^3\bff{r}_i \,G_0(0,\bff{r}_1;E)
W_1(\bff{r}_{1})
 G_0(\bff{r}_1,\bff{r}_2;E) W_2(\bff{r}_{2}) G_0(\bff{r}_2,\bff{r}_3;E)
W_3(\bff{r}_{3}) G_0(\bff{r}_3,0;E)
\nonumber \\ 
&=&\int dr_1 dr_2 dr_3\,
\frac{r_1^{2u_1+1}}{\Gamma(1+2u_1)\cos(\pi
u_1)}\frac{r_2^{2u_2+1}}{\Gamma(1+2u_2)\cos(\pi
u_2)}\frac{r_3^{2u_3+1}}{\Gamma(1+2u_3)\cos(\pi u_3)}
 \nonumber \\
&& \times\bigg\{G_0(0,r_1,E)G_0^{(l=0)}(r_1,r_2,E)G_0^{(l=0)}(r_2,r_3,E)
G_0(0,r_3,E)\bigg\}.
\label{eq:j3} 
\end{eqnarray}
For the zeroth-order $S$-wave Coulomb Green functions we use the 
representations \cite{WiWo61,Voloshin:1985bd} 
\begin{equation}
\label{eq:greenint}
G_0(0,r_i,E)=-\frac{im^2v}{2\pi}e^{imvr_i}\int_{0}^{\infty}dt\,
e^{2imvr_it}\left(\frac{1+t}{t}\right)^{\!\lambda},
\end{equation}
\begin{eqnarray}
\label{eq:greensum} G_0^{(l=0)}(r_i,r_j,E)&=&-\frac{im^2
v}{2\pi}\,e^{imv(r_i+r_j)}\sum_{n=0}^{\infty}
\frac{L_n^{(1)}(-2imvr_i)L_n^{(1)}(-2imvr_j)}{(n+1)(n+1-\lambda)}
\end{eqnarray}
with $v\equiv \sqrt{(E+i\epsilon)/m}$ 
and $\lambda=iC_F\alpha_s/(2v)$. The $L_n^{(l)}(x)$ are the
Laguerre polynomials. The integrals over $r_i$ can now be 
factorized into two functions $H(u,n)$ and $K(u,n,j)$, 
and we obtain 
\begin{eqnarray}
J_{3}&=&\left(\frac{m}{4\pi}\right)^4\sum_{n=0}^{\infty}\sum_{j=0}^{\infty}
\frac{H(u_1,n)K(u_2,n,j)H(u_3,j)}{(n+1)(n+1-\lambda)(j+1)(j+1-\lambda)},
\label{doublesum}
\end{eqnarray}
where (defining $s=-2 i m v r$)
\begin{eqnarray}
H(u,n)&\equiv & \frac{1}{\Gamma(1+2u)\cos\pi
u}\left(\frac{e^{i\pi}}{4m^2v^2}\right)^{\!u}
 \int_0^{\infty}\!dt\left(\frac{1+t}{t}\right)^{\!\lambda}
\int_0^{\infty}\!ds \, e^{-(1+t)s}s^{2u+1}L_n^{(1)}(s),
\phantom{mov}
\\
K(u,n,j)&\equiv & \frac{1}{\Gamma(1+2u)\cos(\pi
u)}\frac{-1}{4m^2v^2}\left(\frac{e^{i\pi}}{4m^2v^2}\right)^{u}
\int_0^{\infty}\!ds\,
s^{2u+1}e^{-s}L_n^{(1)}(s)L_j^{(1)}(s).
\end{eqnarray}

Performing the integrations in $H(u,n)$ (substituting $x=1/(1+t)$ in 
the  $t$-integral), we find  
\begin{eqnarray}
H(u,n)&=&(-4mE)^{-u}\frac{(1+2u)(n+1)}{\cos (\pi u)}
\int_0^{\infty}dt \,
t^{-\lambda}(1+t)^{\lambda-2-2u}\,_2F_1\left(-n,2+2u;2;\frac{1}{1+t}\right)
\nonumber\\
&=& 
\frac{(n+1)\Gamma(1-\lambda)}{\cos(\pi u)(-4mE)^{u}}
\sum_{k=0}^{n}\frac{(-1)^k n!}{k!(n-k)!}
\frac{\Gamma(2+k+2u)\Gamma(1+k+2u)}
{\Gamma(1+2u)\Gamma(k+2)\Gamma(2+k+2u-\lambda)}.
\end{eqnarray}
The sum can be expressed in terms of the hypergeometric function 
$_3F_2(-n,2+2u,1+2u;2,2+2u-\lambda;1)$, but it is simpler to perform
the expansion in the generating variable $u$ directly. We need 
the first two terms in the expansion. With the help of the generating 
sum 
\begin{equation}
\sum_{k=0}^{n}\frac{(-1)^k n!}{k!(n-k)!}
\frac{\Gamma(1+k+a)}{\Gamma(2+k-\lambda)} = 
\frac{\Gamma(1+a)\Gamma(1-a+n-\lambda)}{
\Gamma(2+n-\lambda)\Gamma(1-a-\lambda)}
\end{equation}
we find 
\begin{eqnarray}
 H(0,n) &=& 
\frac{n+1}{n+1-\lambda},
\\
H^{\prime}(0,n) &=&
(n+1)\Gamma(1-\lambda)\sum_{k=0}^{n}\frac{(-1)^k n!}{k!(n-k)!}
\frac{\Gamma(1+k)}{\Gamma(2+k-\lambda)} \bigg\{2\gamma_{E}-\ln(-4mE)
 \nonumber \\
&& +\,2\Psi(1+k)+2\Psi(2+k)-2\Psi(2+k-\lambda)\bigg\}
 \nonumber \\
&=& 
\frac{n+1}{n+1-\lambda}
\bigg\{\!-\ln(-4mE)-2\Big[\gamma_E+\Psi(n+1-\lambda)\Big]
 \nonumber \\ && 
\hspace*{2.1cm}+\,\frac{2\lambda}{(n+1)}\Big[
\Psi(1-\lambda)-\Psi(n+2-\lambda)\Big]\bigg\},
\label{eq:hprime} \end{eqnarray}
where $\Psi(z)$ denotes Euler's Psi-function, and 
$\gamma_E=0.577216\ldots$ is Euler's constant.
This result has already been obtained in \cite{Beneke:1999qg}.

Similarly, for $K(u,n,j)$ we need the expansion up to the first order 
in $u$. Here we obtain 
\begin{eqnarray}
K(0,n,j)&=&-\frac{n+1}{4m^2 v^2}\,\delta_{nj},
\\
K^{\prime}(0,n,j)&=&-\frac{1}{4m^2 v^2} \Big[I + 
(n+1) \,\delta_{nj} \left(2\gamma_E-\ln(-4mE)\right)\Big]
\label{eq:kprime}\end{eqnarray}
with
\begin{eqnarray}
I&=&\int_0^{\infty}ds \,2s\ln(s) e^{-s}L_n^{(1)}(s)L_j^{(1)}(s).
\end{eqnarray}
To solve the integral $I$, the Laguerre polynomials are expressed
through their generating functions. Then with
\begin{eqnarray}
\frac{e^{-\frac{z\,u}{1-u}}}{(1-u)^2}=\sum_{s=0}^{\infty}u^s
L_s^{(1)}(z),
\end{eqnarray}
and 
\begin{eqnarray}
\int_0^{\infty}ds \,2s\ln(s)
e^{-s}\frac{e^{-\frac{s\,v}{1-v}}}{(1-v)^2}
\frac{e^{-\frac{s\,w}{1-w}}}{(1-w)^2}=
-\frac{2\Big[-1+\gamma_E+\ln\left(\frac{1-vw}{(v-1)(w-1)}\right)\Big]}
{(vw-1)^2},
\end{eqnarray}
the integral $I$ is expressed as  
\begin{eqnarray}
I&=&\frac{1}{n!}\frac{1}{j!}\frac{\partial^n}{\partial
v^n}\frac{\partial^j}{\partial
w^j}\left[-\frac{2\left(-1+\gamma_E+
\ln\left(\frac{1-vw}{(v-1)(w-1)}\right)\right)}{(vw-1)^2}\right]_{v=w=0}
\nonumber\\[0.2cm]
&=&\left\{ \begin{array}{cc}
  2+2(1+n)\Psi(1+n) & \hskip0.4cm\textrm{if} \; n=j\\[0.2cm]
  -2\frac{\displaystyle \min (n,j)+1}{\displaystyle |j-n|} & 
  \hskip0.4cm\textrm{if} \; n\neq j \\
\end{array}\right.
\end{eqnarray}
Hence the final result for $K^{\prime}(0,n,j)$ reads 
\begin{eqnarray}
K^{\prime}(0,n,j)&=& -\frac{1}{4m^2 v^2}\left\{ \begin{array}{cc}
  2+(n+1)\Big\{2[\gamma_E+\Psi(1+n)]-\ln(-4mE)\Big\} & 
  \hskip0.4cm\textrm{if} \; n=j\hskip0.8cm\\[0.2cm]
  -2\,\frac{\displaystyle \min (n,j)+1}{\displaystyle |j-n|} & 
  \hskip0.4cm\textrm{if} \; n\neq j \hskip0.8cm
\end{array}\right.
\end{eqnarray}

At this point, we have expressed the threefold
insertion of the perturbation potential in terms of a doubly infinite 
sum involving Euler's Psi-function, see (\ref{doublesum}) . 
These sums converge rapidly when 
the energy argument is evaluated along a line parallel to the real
axis as required in the calculation of the top quark pair production
cross section. In order to obtain the third-order correction to 
the energy levels and wave functions at the origin in 
Section~\ref{sec:result}, we extract analytically the pole part 
of $J_3$ when $E$ approaches the leading-order $S$-wave bound 
state energies $E_n^{(0)}$, which correspond to $\lambda=n$. This 
results in multiple sums, which, after a tedious reduction, can all 
be expressed in terms of zeta-functions and nested harmonic sums.

\end{document}